\documentclass{amsart}
\usepackage{amssymb}
\usepackage{amsfonts}
\usepackage{latexsym}
\newtheorem{definition}{Definition}
\begin{document}
\title{Representation of State Property Systems}
\author{D. Aerts}
\address{Foundations of the Exact Sciences (FUND), Department of Mathematics, Vrije Universiteit Brussel, 1160 Brussels, Belgium}
\email{diraerts@vub.ac.be}
\author{S. Pulmannov\'a}
\address{Mathematical Institute, Slovak Academy of Sciences, SK-814 73 Bratislava, Slovakia}
\email{pulmann@mat.savba.sk}
\keywords{Quantum mechanics, axiomatics, state, property, superposition, representation}
\maketitle
\begin{abstract}
A `state property system' is the mathematical structure which models an arbitrary physical system by means of its set of states, its set of properties, and a relation of `actuality of a certain property for a certain state'. We work out a new axiomatization for standard quantum mechanics, starting with the basic notion of state property system, and making use of a generalization of the standard quantum mechanical notion of `superposition' for state property systems.
\end{abstract}
\section{Introduction}

\noindent
In standard quantum mechanics a state $p_{\bar c}$ of a quantum entity
$S$ is represented by the one dimensional subspace or the ray
$\bar c$ of a separable complex Hilbert space $\mathcal H$.
An experiment $e_A$ testing an observable $A$ is represented by a self
adjoint operator $\mathcal A$ on ${\mathcal H}$, and the set of outcomes of this
experiment $e_A$ is the spectrum $spec({\mathcal A})$ of this self-adjoint operator $\mathcal A$.
Measurable subsets $B
\subset spec({\mathcal A})$ represent the events (in the sense of
probability theory) of outcomes. The interaction of the experiment
$e_A$ with the physical entity being in state $p_{\bar c}$
is described in the following way: (1) the probability for a
specific event $B \subset spec({\mathcal A})$ to occur if the entity is in a
specific state $p_{\bar c}$ is given by
$\langle c, P_B(c) \rangle$, where $P_B$ is the spectral projection
corresponding to $B$,  $c$ is the unit vector in the ray ${\bar c}$ representing state $p_{\bar c}$,
and $\langle \ ,\ \rangle$ is the inproduct in the Hilbert space ${\mathcal H}$ ; (2)
if the outcome is contained in
$B$, the state
$p_{\bar c}$ is changed to $p_{\bar d}$ where ${\bar d}$ is the ray
generated by
$P_B(c)$.

Hence in standard quantum mechanics the states and experiments are represented by means of mathematical entities of a complex
Hilbert space. The crucial role that is played by this complex Hilbert space is very much {\it ad hoc}, in the
sense that there are no physically plausible reasons why
the Hilbert space structure should be at the origin of both the
structure of the state space, as well as the structure of the
experiments. 

This initiated the search for an
axiomatic theory for quantum mechanics where the Hilbert space
structure would be derived from more general and physically more
plausible axioms. The area of forming physical models in the field
of quantum mechanics is very large, and often involves philosophical
problems of physics.  Let us mention some of the most well known
axiomatic approaches: the algebraic approach \cite{segal, emch,
haagkastler}, where the basic notions are observables, the convexity
approach \cite{alfsenschults, edwards1, edwards2, mielnik1,
mielnik2}, where the basic notion is the convex set of states, the
empirical logic approach
\cite{foulisrandall01,randallfoulis01,randallfoulis02,randallfoulis03,randallfoulis04}
where the authors start with primitive notions of an operation or a
test, and the quantum logic approach
\cite{birkhoffvonneumann01,zierler01,mackey01,piron01,jauch01,varadarajan01,
beltrametticassinelli01} which starts with the set of experimental
propositions.

Due to the original focus \cite{birkhoffvonneumann01} on
the collection of `experimental propositions' of a physical entity---with the conviction that such an `experimental proposition' would
be a good basic concept---most of the later axiomatics were
constructed taking as their basic concept the set
${\mathcal L}$ of experimental propositions concerning an entity
$S$. The first breakthrough came with a theorem of
Constantin Piron, who proved that if ${\mathcal L}$ is a complete
[axiom 1], orthocomplemented [axiom 2] atomic [axiom 3] lattice,
which is weakly modular [axiom 4] and satisfies the covering law
[axiom 5], then each irreducible component of the lattice ${\mathcal
L}$ can be represented as the lattice of all `biorthogonal'
subspaces of a vector space $V$ over a division ring $K$ (with some
other properties satisfied that we shall not explicit here) \cite{piron01}. Such a
vector space is called an `orthomodular space' and also sometimes a
`generalized Hilbert space'. It can be proven that an infinite
dimensional orthomodular space over a division ring which is the
real or complex numbers, or the quaternions, is a Hilbert space.
For a long time there did not even exist any other example of an
infinite dimensional orthomodular space. The search for a further
characterization of the real, complex or quaternionic Hilbert space
started \cite{wilbur01}. Then Hans Keller constructed a non classical
orthomodular space \cite{keller01}, and recently Maria Pia Sol{\`e}r proved that any orthomodular space that contains an infinite
orthonormal sequence is a real, complex or quaternionic Hilbert
space \cite{soler01,holland01}. It is under investigation in
which way this result of Sol{\`e}r can be used to formulate new
physically plausible axioms \cite{holland01,pulmannova01,pulmannova02,aertsvansteirteghem01}.

The axiomatic approach, apart from delivering an axiomatic foundations for quantum mechanics, has been used fruitfully to study concrete problems in quantum mechanics. As an example we mention the problem of the description of joint quantum entities, and the problems of entanglement, non-locality and appearance of the complex numbers in quantum mechanics. Most recently this problem has been studied within the axiomatic approach with very interesting results \cite{ischi01,ischi02,ischi03,ischi04,watanabe01,watanabe02,niestegge01}.

Next to the idea of finding axioms that introduce the Hilbert space structure step by step, was the attempt of founding the basic notions for this axiomatics in a physically clear and operational way. `Operationality' means that the axioms
should be introduced in such a way that they can be related   to `real physical operations' that can be performed in the laboratory.

The approaches that have tried to formulate quantum mechanics
operationally are, the Geneva-Brussels approach \cite{piron01,jauch01,piron02,aerts01,aerts02,aerts03,aerts04,piron03,piron04}, the Amherst
approach \cite{foulisrandall01,randallfoulis01,randallfoulis02,randallfoulis03,randallfoulis04}, and the Marburg
approach \cite{ludwig01,ludwig02}. In the present article we elaborate further on the Geneva-Brussels approach. Already in the last versions of the formalism that were presented in this approach the power of making a good distinction between the
mathematical aspects of the formalism and its physical foundations had been identified \cite{aerts05,aerts06}. Let us explain more concretely what we
mean. In the older founding papers of the Geneva-Brussels approach
\cite{piron02,aerts01,aerts02,aerts03,aerts04,piron03,piron04}, although the physical foundation of the
formalism is defined in a clear way, and the resulting mathematical structures are treated rigorously, it is not always
clear what are the `purely mathematical' properties of the structures that are at the origin of the results. That is the reason that
in more recent work on the formalism we have made an attempt to divide up the physical foundation and the resulting mathematical
structure as much as possible. We first explain in which way certain aspects of the mathematical structure
arise from the physical foundation, but then, in a second step, define these aspects in a strictly mathematical way, such that
propositions and theorems can be proven, `only' using the mathematical structure without physical interpretation. Afterwards, the results of
these propositions and theorems can then be interpreted in a physical way again. This not only opens the way for mathematicians to start
working on the structures, but also lends a greater axiomatic strength to the whole approach on the fundamental level. More concretely, it is the
mathematical structure of a `state property system' that has been identified to be the proper mathematical structure of the Geneva-Brussels approach, {\it i.e.} the structure used to describe a physical entity by means of its states and
properties \cite{aerts05,aerts06,aertsetal01}. This step turned out to be fruitful from the start, since we could prove that
a state property system as a mathematical structure is isomorphic to a closure space \cite{aerts05,aerts06,aertsetal01}. This
means that the mathematics of closure spaces can be translated to the mathematics of state property systems, and in this sense
becomes relevant for the foundations of quantum mechanics. The step of dividing up the mathematics from the physics in a
systematic way also led to a scheme to derive the morphisms for the structures that we consider from a covariance principle
rooted in the relation of a subentity to the entity of which it is a subentity \cite{aerts06,aertsetal01}. This paved the
way to a categorical study of the mathematical structures involved.

Not only was it possible to connect with a state property system a closure space in an isomorphic way, but, after we had
introduced the morphisms starting from a merological covariance principle, it was possible to prove that the category of state
property systems and their morphisms, that we have named {\bf SP}, is equivalent with the category of closure
spaces and continuous functions, denoted by {\bf Cls} \cite{aerts06,aertsetal01}. More specifically we could prove that {\bf
SP} is the amnestic modification of {\bf Cls} \cite{aertsetal02}.

It could be proven that some of the axioms of axiomatic quantum
mechanics \cite{piron01,aerts01,aerts02} correspond to separation properties of the
corresponding closure spaces \cite{vansteirteghem01}. More concretely, the axiom of state
determination in a state property system \cite{aerts05} is equivalent with the
$T_0$ separation axiom of the corresponding closure space \cite{vansteirteghem01,vansteirteghem02}, and the axiom of
atomicity in a state property system \cite{aerts05} is equivalent with the $T_1$ separation axiom of the corresponding
closure space
\cite{vandervoorde01,vandervoorde02}. More recently it has been shown
that `classical properties' \cite{aerts01,aerts03,piron03,piron04} of the state property system correspond to
clopen (open and closed) sets of the closure space \cite{aertsetal03,aertsetal04,aertsetal05}, and, explicitly making use of the categorical
equivalence, a decomposition theorem for a state property system into its nonclassical components can be proved that
corresponds to the decomposition of the corresponding closure space into its connected components
\cite{aertsetal03,aertsetal04,aertsetal05}.

 In the present article we put forward a new axiomatization for standard
 quantum mechanics, starting with the basic notion of `state property system',
 and founded on the concept of `superposition', that started in the
 quantum logic approach and was developed in \cite{PtPu}.
 The general reason for introducing the new axiomatization is to put under
 one roof the Geneva-Brussels approach in its recent form and the
 quantum logic approach and to combine the algebraic approach
 and the probabilistic approach. We also wish to find out how the recent
 development in projective geometry (see \cite{FF}) can be
 reflected in the axiomatization. A more specific reason for this new axiomatization is to take it as a mathematical basis for further research into the problem of the description of joint quantum systems. Both authors have done extensive research on the problem of the description of joint quantum systems \cite{aertsetal05,aertsetal06,aertsetal07,aerts07,aerts02,pulmannova03,aerts08,aerts09,pulmannova04,pulmannova05}. One of hard the problems is that all type of product constructions on the level of the quantum logic structure give rise to a situation where the joint quantum entity only has product states of the subentities. On the level of the Hilbert space, the joint system of two quantum systems is described by means of the tensor product of the Hilbert spaces of the subsystems, and in this case there is an abundance of non-product states, giving rise to the well known phenomenon of quantum entanglement.
We plan to study the still open problem of coupled physical systems with entanglement, by investigating in which way we can introduce `superpositions between product states' by means of the notion of `superposition' which we introduced in this axiomatization on the level of the quantum logic.

The notion of a `superposition of states'  was introduced by
Varadarajan \cite{Var} for states as probability measures on
quantum logics, i.e., orthomodular lattices. In the same sense it is
also used in \cite{PtPu}. In the present paper, we use a more
general frame of a state property system to introduce the notion of
`superposition'. We use  superpositions to create two kinds of
closure operations. The first one, together with a few simple
additional axioms, enables us to associate the structure of a
projective geometry with our state property system. A very useful
tool here is the material presented in the recent book by Faure and Fr\"olicher \cite{FF}. The
first of our closure operations based on superposition leads to
the formation of subspaces of a projective geometry. The second of
our closure operations enables us to characterize closed subspaces
of the projective geometry. Probabilities enter into play in order
to introduce orthocomplementation on a subset ${\mathcal L}_0$ of
the lattice ${\mathcal L}$, and we show that ${\mathcal L}_0$ can be
organized into a $\sigma$- orthomodular poset with an order
determining set of probability measures, which are supported by
elements of the set $\Sigma$. The set ${\mathcal L}_0$ may be
interpreted as a set of measurable properties, and may depend on the
present state of knowledge and experimental techniques. In the
following parts of the article, conditions are found under which the
orthocomplementation can be extended to the whole ${\mathcal L}$,
and ${\mathcal L}$ then becomes a complete, atomistic,
orthocomplemented lattice. Moreover, ${\mathcal L}$ can be related
with the closed subspaces of the projective geometry via the so-called
Cartan map. The notion of `superposition principle' is introduced
to obtain irreducibility of the projective geometry. More generally,
sectors are introduced as the minimal subspaces in which the
superposition principle holds, and their topological
characterization as clopen subspaces is derived. In the following the classical properties (or the
superselection rules) are specified, and it is shown that they
correspond exactly to the central elements of the lattice ${\mathcal
L}$. In the following, we study conditions under which our
projective geometry may admit some deeper properties, described in \cite{FF}, such as the Mackey property or to become an
orthogeometry. Although not all of our axioms have a physical
meaning, we try to specify simple axioms which enable us
to obtain different stages of the projective geometry. Eventually we
find conditions under which a vector space can be associated with
our state property system, and we finish with an analogue of the
famous Piron theorem.

\section{State Property Systems and Superposition}
\begin{definition}\label{de:1}{\rm \cite{aerts05,aerts06,aertsetal01}}
We say that $(\Sigma,{\mathcal L},\xi)$ is a state-property system
if $(\Sigma,<)$ is a pre-ordered set, $({\mathcal L},<,\wedge,
\vee)$ is a complete lattice with the greatest element $I$ and the
smallest element $0$, and $\xi$ is a function
\begin{equation}\label{eq:1}
\xi :\Sigma \to {\mathcal P}({\mathcal L})
\end{equation}
such that for $p \in \Sigma$ and $(a_i)_i\subseteq {\mathcal L}$,
we have
\begin{eqnarray}
& I\in \xi(p),\\
& 0\notin \xi(p),\\
& a_i\in \xi(p)\ \forall i \Leftrightarrow  \wedge_i a_i \in
\xi(p)\, \mbox{(for an arbitrary set of indices)}
\end{eqnarray}
and for $p,q\in \Sigma$ and $a,b\in {\mathcal L}$ we have
\begin{eqnarray}
& p<q \Leftrightarrow \xi(q)\subseteq \xi(p)\\
& a\leq b \Leftrightarrow \forall r\in \Sigma : a\in \xi(r)
\Rightarrow b\in \xi(r)
\end{eqnarray}
\end{definition}
Elements of $\Sigma$ are called {\it states}, elements of
$\mathcal L$ are called {\it properties}.

Let $(\Sigma, {\mathcal L}, \xi)$ be a state-property system. For
$S\subset \Sigma$ define $S\mapsto \lambda(S)$ as follows. First
define, for any $p,q\in \Sigma$,

\begin{equation}\label{eq:2}
\lambda \{ p,q\}:=\{ s\in \Sigma :a\in \xi(p)\cap \xi(q)
\Rightarrow a\in \xi(s)\}.
\end{equation}
We will say that a subset $S\subseteq \Sigma$ is {\it
$\lambda$-closed} if  for any $p,q\in S$ we have $\lambda \{ p,q\}
\subset S$.  Denote by ${\mathcal L}(\Sigma)$ the set of all
$\lambda$-closed subsets. For any $P\subset \Sigma$, define
\begin{equation}\label{eq:3}
\lambda(P):=\bigcap \{ G:G\in {\mathcal L}(\Sigma), P\subset G\}.
\end{equation}
That is, $\lambda(P)$ is the intersection of all $\lambda$-closed
subsets of $\Sigma$ that contain $P$.

\newtheorem{lemma}{Lemma}
\begin{lemma}\label{le:1} {\rm (i)} For every subset $P\subseteq
\Sigma$, $\lambda(P)\in {\mathcal L}(\Sigma)$.  {\rm (ii)} A
subset $S\subseteq \Sigma$ is $\lambda$-closed if and only if
$S=\lambda(S)$.
\end{lemma}
\begin{proof} (i) Let $p,q\in \lambda(P)$, then $p,q\in G$ for every
$P\subseteq G\in {\mathcal L}(\Sigma)$. Therefore $\lambda
\{p,q\}\subseteq G$ for every such $G$, and consequently
$\lambda\{ p,q\}\subseteq \lambda(P)$.

(ii) If $S=\lambda(S)$, then $S\in {\mathcal L}(\Sigma)$ by (i).
If $S$ is $\lambda$-closed, then clearly, $S$ is the smallest
$\lambda$-closed subset of $\Sigma$ containing $S$, hence
$\lambda(S)=S$.
\end{proof}

That is,
\begin{equation}\label{eq:L}
 {\mathcal L}(\Sigma)=\{ S\subset \Sigma : S=\lambda (S)\}.
\end{equation}

Proof of the following statement is immediate.
\begin{lemma}\label{le:2}
The mapping $\lambda :P\mapsto \lambda(P)$ satisfies the following
properties:
\begin{enumerate}
\item[{\rm(C1)}] $P\subseteq \lambda(P)$,
\item[{\rm(C2)}] $P_1\subseteq \lambda(P_2)$ $\Rightarrow$
$\lambda(P_1)\subseteq \lambda(P_2)$.
\end{enumerate}
\end{lemma}

We recall that a map $C:{\mathcal P}(X)\to {\mathcal P}(X)$
satisfying conditions (C1) and (C2) is a {\it closure operator} on
the set $X$ (\cite[Def. 3.1.1]{FF}). Consequently, $\lambda$ is a
closure operator on the set $\Sigma$.

According to \cite[Remark 3.1.2]{FF}, the following conditions are
satisfied.
\begin{enumerate}
\item[($1^0$)] $\lambda(A\cup B)=\lambda(\lambda(A)\cup
B)=\lambda(\lambda(A)\cup \lambda(B))$,
\item[($2^0$)] $\lambda(A\cap B)\subseteq \lambda(\lambda(A)\cap
B)\subseteq \lambda(A)\cap \lambda(B)$.
\end{enumerate}
More generally,
\begin{enumerate}
\item[($3^0$)] $\lambda(\bigcup{\mathcal A})=\lambda(\bigcup
\lambda({\mathcal A}))$,
\item[($4^0)$] $\lambda(\bigcap{\mathcal A})
\subseteq \bigcap \lambda({\mathcal A})$,
\end{enumerate}
where ${\mathcal A} \subseteq {\mathcal P}(\Sigma)$ is an
arbitrary subset and $\lambda({\mathcal A})$ denotes the set $\{
\lambda(A) : A\in {\mathcal A}\}$.

Recall that a system ${\mathcal S}$ of sets is an {\it
intersection system} if ${\mathcal A}\subseteq {\mathcal S}$
implies $\bigcap {\mathcal A}\in {\mathcal S}$. By \cite[Prop.
3.1.4]{FF}, ${\mathcal L}(\Sigma)$ is an intersection system.

\begin{definition}\label{de:2} We say that a state-property system
$(\Sigma, {\mathcal L}, \xi)$ satisfies property
\begin{enumerate}
\item[{\rm(A)}] if there are at least two distinct states $r,s\in \Sigma$
and for all $p,q\in \Sigma$, $\xi(p)\subset \xi(q) \Rightarrow
p=q$.
\end{enumerate}
\end{definition}

Property (A) implies that $\xi:\Sigma \to {\mathcal P}({\mathcal
L})$ is injective. By \cite[Proposition 14]{aertsetal01}, the
pre-order $<$ on $\Sigma$ defined by (5) of Definition \ref{de:1} is
a partial order. Property (A) implies that $\Sigma$ has only a
trivial order $p < q$ iff $p=q$.

Recall that a closure operator $C$ on a set $X$ is called {\it
simple} if it satisfies the additional axiom:
\begin{enumerate}
\item[(C5)] $C(\emptyset)=\emptyset$ and $C(x)=\{ x\}$ for every
$x\in X$.
\end{enumerate}
(We write $C(x)$ instead of $C(\{x\})$). If $X$ contains at least
two different points, then the second property in (C5) implies the
first one. Indeed, $\emptyset \subset \{ x\}, \emptyset \subset \{
y\}$ implies $C(\emptyset)\subset C(x)\cap C(y)=\{ x\}\cap \{
y\}=\emptyset$.

\begin{lemma}\label{le:3} Let $(\Sigma,{\mathcal L},\xi)$ be a
state-property system such that there are at least two different
states $p,q\in \Sigma$. Then the closure operator $\lambda$ is
simple if and only if property $(A)$ of Definition \ref{de:2}
holds.
\end{lemma}
\begin{proof}
If (A) holds, then for every $p\in \Sigma$,
\begin{eqnarray*}
\lambda \{ p\}& =\{ s\in \Sigma :a\in \xi(p) \Rightarrow a\in \xi(s)\}\\
& = \{ s\in \Sigma : \xi(p) \subset \xi(s)\} =\{ p\}.
\end{eqnarray*}
If $\lambda(\emptyset)$ contains $r$, then
$\lambda(\emptyset)\subseteq \lambda(q)=\{ q\}$ implies $r\in \{
q\}$, hence $r=q$ for all $q$, a contradiction.

Conversely, if $\lambda$ is simple and $\xi(p)\subseteq \xi(q)$
for $p\neq q$, then
$$
\lambda(p)=\{ s\in \Sigma : \xi(p)\subseteq \xi(s)\}\ni q
$$
contradicting $\lambda(p)=\{ p\}$.
\end{proof}

\newtheorem{proposition}{Proposition}

\begin{proposition}\label{pr:1} Let $(\Sigma, {\mathcal L},\xi)$ be
a state property system satisfying (A). Then  ${\mathcal L}$ is a
complete atomistic lattice with the lattice operations
$$
\bigwedge S_{\alpha}=\bigcap S_{\alpha}, \, \bigvee
S_{\alpha}=\lambda(\bigcup S_{\alpha}).
$$
\end{proposition}
\begin{proof} Follows by \cite[prop. 3.1.4]{FF}.
\end{proof}

In what follows, we introduce the notion of a superposition of states in analogy with \cite{varadarajan01}.

\begin{definition}\label{de:3} A state $p\in \Sigma$ is a
superposition of a set of states $S$, $S\subseteq \Sigma$, if for
each $a\in {\mathcal L}$, $a\in \xi(s)$ for all $s\in S$ implies
$a\in \xi(p)$, i.e. if $\bigcap_{s\in S}\xi(s)\subseteq \xi(p)$.

For $S\subseteq \Sigma$, define
\begin{equation}
{\bar S}=\{ p\in \Sigma : \bigcap_{s\in S} \xi(s)\subseteq
\xi(p)\}.
\end{equation}
\end{definition}
That is, ${\bar S}$ is the set of all superpositions of states in
$S$. Obviously, for arbitrary $p,q\in \Sigma$,
\begin{equation}
\{ p,q\}^- = \lambda \{ p,q\}.
\end{equation}

\begin{definition}\label{de:4} A state $p\in \Sigma$ is a minimal
superposition of a subset $S\subseteq \Sigma$ if
\begin{enumerate}
\item[\rm{ (i)}] $p\in {\bar S}$,
\item[\rm{ (ii)}] $p\notin {\bar Q}$ for any proper subset
$Q\subseteq S$.
\end{enumerate}
\end{definition}

For example, if (A) holds, then $s\in \{ p,q\}^-$ is a minimal
superposition iff $s\neq p$, $s\neq q$.

\begin{definition}\label{de:5} Let $(\Sigma, {\mathcal L}, \xi)$ be a
state-property system.

{\rm (1)} We will say that a minimal superposition principle {\rm
(MSP)} holds for $(\Sigma, {\mathcal L}, \xi)$ if for every subset
$S\subseteq \Sigma$ and for every minimal superposition $p$ of
$S$,
\begin{equation}\label{eq:MSP}
\{ S_1\cup p\}^- \cap {\bar S_2} \neq \emptyset
\end{equation}
whenever $S_1,S_2$ are proper subsets of $S$ such that $S_1\cap
S_2=\emptyset$ and $S_1\cup S_2=S$.

{\rm (2)} We will say that a finite MSP  {\rm (f-MSP)} holds for
$(\Sigma, {\mathcal L}, \xi)$ if {\rm (\ref{eq:MSP})} holds for
every finite subset $S\subseteq \Sigma$.

{\rm (3)} We will say that an $n$-MSP holds for $(\Sigma,
{\mathcal L},\xi)$ if {\rm (\ref{eq:MSP})} holds for every subset
$S\subseteq \Sigma$ with the cardinality at most $n$.
\end{definition}

For example,the 3-MSP holds for a state-property system $(\Sigma,
{\mathcal L},\xi)$ iff for every $p,q,r,s\in \Sigma$ (not
necessarily all different), if $p\in \{ q,r,s\}^-$ is a minimal
superposition, then $\{ p,q\}^-\cap \{ r,s\}^-\neq \emptyset$ and
$\{ p,q,r\}^-\cap \{ s\}^-\neq \emptyset$. Clearly,  MSP implies
f-MSP, which in turn implies n-MSP for every $n\in \mathbb N$.
Observe also that if (A) holds, then 2-MSP is the following
exchange property : if $r\in \{ p,q\}^-$ and $r\neq p,q$, then
$p\in \{ r,q\}^-$.
\newtheorem{theorem}{Theorem}

\begin{theorem}\label{th:1} Let $(\Sigma,{\mathcal L},\xi)$ be a
state-property system with properties  {\rm(A)} and {\rm 3-MSP}.
Then the operator $*:\Sigma \times \Sigma \to {\mathcal
P}(\Sigma)$ defined by $p*q=\lambda \{ p,q\}$ has the following
properties:
\begin{enumerate}
\item[{\rm (P1)}] $p*p=\{ p\}$,
\item[{\rm(P2)}] $p\in p*q$ for all $p,q\in \Sigma$,
\item[{\rm(P3)}] $p\in q*r$ and $r\in s*t$ and $p\neq s$ imply
$(p*s)\cap(q*t)\neq \emptyset$
\end{enumerate}
That is, the system $(\Sigma,*)$ is a projective geometry
\cite{FF}.
\end{theorem}
\begin{proof} (P1) By (A), $\lambda\{ p,p\}=\{ p\}$.

(P2) Clearly, $p*q=\lambda\{p,q\}\supseteq \{ p\}$.

(P3) From $p\in \{ q,r\}^-$ and $r\in \{ s,t\}^-$ we obtain $p\in
\{ q,s,t\}^-$. If $p\in q*s$ resp.  $p\in s*t$, then either $p=q$,
resp. $p=t$, or 2-MSP implies that $q\in p*s$, resp. $t\in p*s$.
In every case, (P3) is satisfied. It remain the case that either
$p\in q*t$ or $p$ is a minimal superposition of $\{ q,s,t\}$. In
the first case, $p\in (p*s)\cap(q*t)$ holds by (P2). In the second
case the statement follows by 3-MSP.
\end{proof}

By \cite[Def. 2.3.1]{FF}, the $\lambda$-closed subsets of $\Sigma$
coincide, under the suppositions of Theorem \ref{th:1}, with the
subspaces of the projective geometry $(\Sigma,*)$ associated with
$(\Sigma,{\mathcal L},\xi)$. Consequently, we have the following.

\begin{theorem}\label{th:2} Let $(\Sigma, {\mathcal L},\xi)$ be a
state-property system satisfying conditions {\rm(A)} and {\rm
3-MSP}. Then ${\mathcal L}(\Sigma)$ is a projective lattice,i.e.,
a complete atomistic meet-continuous modular lattice.
\end{theorem}

Therefore we will call the elements of ${\mathcal L}(\Sigma)$ the
{\it subspaces} of $\Sigma$. From the next theorem we can derive
what properties satisfies the closure operator $\lambda$ on the
system $(\Sigma, *)$ with $p*q=\lambda \{ p,q\}$. (We write
$\lambda(x)$ instead of $\lambda (\{ x\})$ and $\lambda(A\cup x)$
instead of $\lambda(A\cup \{ x\})$.)

\begin{theorem}\label{th:3}
Let $(\Sigma,{\mathcal L}, \xi)$ be a state-property system such
that $(\Sigma, *)$ with $p*q=\lambda \{ p,q \}$ is a projective
geometry, i.e., properties {\rm (P1), (P2)} and {\rm (P3)} are
satisfied. Then the closure operator $\lambda$ satisfies the
following conditions.
\begin{enumerate}
\item[{\rm (C3)}] $x\in \lambda(A)$ implies $x\in \lambda(B)$ for
some finite subset $B\subseteq A$, i.e., $\lambda$ is finitary.
\item[{\rm(C4)}] $x\notin \lambda(A)$ and $x\in \lambda(A \cup y)$
imply $y\in \lambda(A\cup x)$, i.e., $\lambda$ satisfies the
exchange property.
\item[{\rm(C5)}] $\lambda(\emptyset)=\emptyset$ and $\lambda(x)=\{
x\}$, i.e., $\lambda$ is simple.
\item[{\rm(C6)}] $\lambda(A\cup B)=\bigcup \{ \lambda \{
x,y\}:x\in \lambda(A)\ \mbox{and} \ y\in \lambda(B)\}$ for every
$A,B\subseteq \Sigma$.
\end{enumerate}
\end{theorem}
\begin{proof} See \cite[Th. 3.3.4]{FF}.
\end{proof}

Notice that (C6) is called a {\it projective law}. According to
\cite[Lemma 3.3.2]{FF}, if a closure operator $C: {\mathcal
P}(X)\to {\mathcal P}(X)$ satisfies the projective law, then for
any nonempty subset $A\subseteq X$ and any $b\in X$ one has
\begin{enumerate}
\item[(C7)] $C(A\cup b)=\bigcup \{ C(x,b):x\in C(A)\}$.
\end{enumerate}
Moreover, the converse holds  provided the closure operator $C$
satisfies (C3), i.e. is finitary. The following proposition
follows by \cite[Proposition 3.3.4]{FF}.

\begin{proposition}\label{pr:proj} Let the closure operator
$\lambda :{\mathcal P}(\Sigma)\to {\mathcal P}(\Sigma)$ satisfy
properties {\rm (C4), (C5)} and {\rm (C7)}. The couple $(\Sigma,
*)$ where $p*q=\lambda \{p,q\}$ is a projective geometry.
\end{proposition}

Now we will study connections between the mappings $A\mapsto
\lambda(A)$ and $A\mapsto {\bar A}$, $A\subseteq \Sigma$. First we
prove the following properties of $A\mapsto {\bar A}$.

\begin{lemma}\label{le:4} Let $(\Sigma, {\mathcal A},\xi)$ be a
state-property system. The operator $A\mapsto {\bar A}$ satisfies
the following properties for every $A,B\subseteq \Sigma$.
\begin{enumerate}
\item[{\rm (i)}] $A\subseteq {\bar A}$.
\item[{\rm(ii)}] $A\subseteq {\bar B} \ \Rightarrow \ {\bar A}\subseteq
{\bar B}$.
\item[{\rm(iii)}] ${\bar A}\in {\mathcal L}(\Sigma)$.
\end{enumerate}
\end{lemma}
\begin{proof} (i) Follows directly from the definition.

(ii) Let $p\in {\bar A}$, i.e., $\bigcap_{s\in A} \xi(s)\subseteq
\xi(p)$. $A\subseteq {\bar B}$ implies that for every $s\in A$,
$\bigcap_{q\in B}\xi(q) \subseteq \xi(s)$, so that $\bigcap_{q\in
B}\xi(q)\subseteq \bigcap_{s\in A}\xi(s)\subseteq \xi(p)$. Hence
$p\in {\bar B}$, and so ${\bar A}\subseteq {\bar B}$.

(iii) From $\lambda \{ p,q\}=\{ p,q\}^-\subseteq {\bar A}$ for
every $p,q\in A$ we see that ${\bar A}$ is $\lambda$-closed, i.e.,
${\bar A}=\lambda({\bar A})$.
\end{proof}

Observe that (i) and (ii) in Lemma \ref{le:4} coincide with the
properties (C1) and (C2), respectively, so that $A\mapsto {\bar
A}$ is a closure operator. Let us denote by ${\mathcal F}(\Sigma)$
the set of superposition-closed subsets of $\Sigma$, that is,
\begin{equation}\label{eq:fsigma}
{\mathcal F}(\Sigma):=\{ S\subset \Sigma :{\bar S}=S\}.
\end{equation}

\begin{proposition}\label{pr:fs} Let $(\Sigma, {\mathcal L},\xi)$
be a state property system satisfying condition {\rm(A)}. Then the
set ${\mathcal F}(\Sigma)$ is a complete atomistic lattice.
Moreover, if $S_i\in {\mathcal F}(\Sigma), i\in I$, for any index
set $I$, then $\bigwedge_{i\in I} S_i=\bigcap_{i\in I} S_i$, and
$\bigvee_{i\in I} S_i=(\bigcup_{i\in I} S_i)^-$.
\end{proposition}
\begin{proof} If $S\subset \Sigma$, $S=\emptyset$ or $S=\{ s\}$,
then $S={\bar S}$ by condition (A). So one-element sets belong to
${\mathcal F}(\Sigma)$ which are atoms in ${\mathcal F}(\Sigma)$.
>From the properties of closure operators (\cite[Prop. 3.1.4]{FF}),
we get $\bigwedge_{i\in I} S_i=\bigcap_{i\in I} S_i$, and
$\bigvee_{i\in I} S_i=(\bigcup_{i\in I} S_i)^-$.
\end{proof}

\begin{theorem}\label{th:4} Let $(\Sigma, {\mathcal L},\xi)$ be a
state-property system such that condition (A) is satisfied.
\begin{enumerate}
\item[{\rm(i)}] If {\rm 3-MSP} holds, then for every $p,q,s\in \Sigma$,
\begin{equation} \label{eq:3-MSP}
\{p,q,s\}^-=\lambda \{ p,q,s\}
\end{equation}
\item[{\rm(ii)}] If {\rm f-MSP} holds, then for every finite subset
$A=\{ s_1,s_2,\ldots, s_n\}\subseteq \Sigma$,
\begin{equation}\label{f-MSP}
\lambda(A)={\bar A}.
\end{equation}
\end{enumerate}
\end{theorem}
\begin{proof} (i) For every $p,q\in \Sigma$, $\lambda \{ p,q\}=\{
p,q\}^-$, and by Lemma \ref{le:4} (iii), $\lambda \{
p,q,s\}\subseteq \{ p,q,s\}^-$ for every $p,q,s\in \Sigma$. To
prove the converse inclusion, let $t\in \{ p,q,s\}^-$. If $t\in \{
p,q\}^-$, then $t\in \lambda \{ p,q\}\subseteq \lambda \{
p,q,s\}$. Hence we may assume that $t$ is a minimal superposition.
Then by 3-MSP, there is $r\in \{ p,t\}^-\cap \{ q,s\}^-$. By
2-MSP, $t\in \lambda \{ r,p\}\subseteq \lambda \{ p,q,s\}$. This
implies that $\{ p,q,s\}^- \subseteq \lambda \{ p,q,s\}$.

(ii) We will proceed by induction. For n=2, the statement holds.
Assume that the statement holds for every $k\leq n$, $k,n\in
\mathbb N$. Let $A=\{ s_1,s_2\ldots,s_n,s_{n+1}\}$, and assume
that $t\in {\bar A}$ is a minimal superposition. By f-MSP and
induction hypothesis,  there is $r\in \{ t,s_{n+1}\}^-\cap \{ s_1,
\ldots, s_n\}^-\subseteq \lambda \{ s_1,\ldots,s_n\}$. Now $t\in
\lambda \{ r, s_{n+1}\}\subseteq \lambda ((\lambda \{
s_1,\ldots,s_n\}\cup s_{n+1})\subseteq \lambda(A)$. If $t$ is not
a minimal superposition, there is a subset $B\subseteq A$ such
that $t\in {\bar B}=\lambda(B)\subseteq \lambda(A)$ by induction
hypothesis. Hence ${\bar A}\subseteq \lambda(A)$. The converse
holds by Lemma \ref{le:4} (iii).
\end{proof}

\section{Probability measures and orthocomplementation}
Let $(\Sigma,{\mathcal L},\xi)$ be a state-property system. Let
there be a subset  ${\mathcal L}_0 \subset {\mathcal L}$ such that
${\mathcal L}_0$ contains $0$ and $1$, and let there be a mapping
$\mu :\Sigma \times{\mathcal L}_0 \to [0,1]$,  $(p,a)\mapsto
\mu_p(a)$, where $[0,1]$ is the unit interval of the reals,  such
that
\begin{enumerate}
\item[{\rm (Oi)}]
$\mu_p(a)=1$ iff $a\in \xi(p)$ ($a\in{\mathcal L}_0$),
\item[{\rm (Oii)}] $a\leq b$ implies $\mu_p(a)\leq \mu_p(b)$
($a,b\in {\mathcal L}_0$),
\item[{\rm (Oiii)}] If $(a_i)_{i=1}^{\infty}\subset {\mathcal L}_0$
is a sequence such that for all $i,j$, and every $p\in \Sigma$,
$$
\mu_p(a_i)+\mu_p(a_j)\leq 1,
$$
then there is  $b\in {\mathcal L}_0$ such that
$$
\mu_p(b)+\sum_{i=1}^{\infty} \mu_p(a_i)=1.
$$
\end{enumerate}
Clearly, $\mu_p(I)=1$ and $\mu_p(0)=0$ for all $p\in \Sigma$.
Define a relation $\perp\subset {\mathcal L}_0\times {\mathcal
L}_0$ by setting $a\perp b$ iff $\mu_p(a)+\mu_p(b)\leq 1$ for all
$p\in \Sigma$. We will say that $a$ and $b$ are {\it orthogonal}
if $a\perp b$.

\begin{lemma}\label{le:5} Let $({\mathcal L}, \Sigma, \xi)$ be a
state-property system. Let ${\mathcal L}_0\subset {\mathcal L}$
and $\mu :\Sigma \times {\mathcal L}_0 \to [0,1]$ satisfy the
assumptions {\rm (Oi) - (Oiii)}. Then
\begin{enumerate}
\item[{\rm(i)}] $\mu_p(a)\leq \mu_p(b)$ for every $p\in \Sigma$
implies $a\leq b$.
\item[{\rm(ii)}] $\mu_p(a)=\mu_p(b)$ for all $p\in \Sigma$ if
and only if $a=b$.
\item[{\rm(iii)}] For every $a\in {\mathcal L}_0$ there is a
unique element $a'\in {\mathcal L}_0$ such that
$\mu_p(a)+\mu_p(a')=1$ for all $p\in \Sigma$. Moreover, the
mapping $a\mapsto a'$ is an orthocomplementation in ${\mathcal
L}_0$, i.e., {\rm(1)} $a\leq b \ \Rightarrow \ b'\leq a'$, {\rm
(2)} $a'':=(a')'=a$, {\rm(3)} $a\vee_0 a'=I$, $a\wedge_0 a'=0$,
where $\vee_0$ and $\wedge _0$ denote the supremum and infimum in
${\mathcal L}_0$, respectively.
\item[{\rm(iv)}] For every sequence $(a_i)_{i=1}^{\infty}$ of
mutually orthogonal elements in ${\mathcal L}_0$, their supremum
$a={\bigvee_0}_{i=1}^{\infty} a_i$ exists in ${\mathcal L}_0$, and
coincides with the supremum of $(a_i)_{i=1}^{\infty}$ in
${\mathcal L}$.
\end{enumerate}
\end{lemma}
\begin{proof} (i) If $\mu_p(a)\leq \mu_p(b)$ for every $p\in \Sigma$, then
$\mu_p(a)=1 \Rightarrow \mu_p(b)=1$, hence by (Oi), $a\in \xi(p)
\Rightarrow b\in \xi(p)$, which implies $a\leq b$.

(ii) follows by (Oii) and (i).

(iii) Let $a\in {\mathcal L}_0$, and consider the sequence
$(a_i)_{i=1}^{\infty}$ where $a_1=a$, $a_i=0$, $i=2,3,\ldots$. By
(Oiii), there is $b\in {\mathcal L}_0$ such that
$\mu_p(b)+\mu_p(a)=1$ for all $p\in \Sigma$, i.e.,
$\mu_p(b)=1-\mu_p(a)$ for all $p\in \Sigma$. Hence we may put
$a'=b$. By (ii), $a'$ is uniquely defined. Now we prove that
$a\mapsto a'$ is an orthocomplementation.

(1) $a\leq b \Rightarrow \mu_p(a)\leq \mu_p(b)$ for all $p\in
\Sigma$, which implies $\mu_p(b')=1-\mu_p(b)\leq
1-\mu_p(a)=\mu_p(a')$ for all $p\in \Sigma$, which by (ii) entails
$b'\leq a'$.

(2) $\mu_p((a')')=1-\mu_p(a')=1-(1-\mu_p(a))=\mu_p(a)$ for all
$p\in \Sigma$, which entails $a''=a$.

(3) Let $c\in {\mathcal L}_0$ be such that $a\geq c, a'\geq c$.
>From $\mu_p(a)+\mu_p(a')=1$ and $\mu_p(a)=1$ iff $a\in \xi(p)$, it
follows that $a\in \xi(p) \Rightarrow a'\notin \xi(p)$, and vice
versa. Hence $a\in \xi(p)$ and $a'\in \xi(p)$ happens for no $p\in
\Sigma$, which entails, by Definition \ref{de:1}, that $a\wedge
a'=0$ in ${\mathcal L}$. Since $0\in {\mathcal L}_0$, the infimum
of $a$ and $a'$ in ${\mathcal L}_0$ is $0$.

Properties (1) and (2) imply de Morgan laws in ${\mathcal L}_0$:
 $a\vee_0 b$ exists, then $(a\vee_0 b)'=a'\wedge_0 b'$, and
$(a\wedge_0 b)'=a'\vee_0 b'$ in the sense that if one side exists,
so does the other, and they are equal. Therefore for every $a\in
{\mathcal L}_0$, $a'\wedge_0 a''=0$ implies $(a'\wedge_0
a'')'=a\vee_0 a'=I$.

(iv) Let $(a_i)_{i=1}^{\infty}$ be a sequence of pairwise
orthogonal elements of ${\mathcal L}_0$. Let $b$ be the element
from (Oiii). Put $a:=b'$, then for every $p\in \Sigma$ we have
$\mu_p(a)=\sum_{i=1}^{\infty} \mu_p(a_i)$. It follows that
$\mu_p(a_i)\leq \mu_p(a)$ for all $i\in \mathbb N$, and for all
$p\in \Sigma$. Hence $a$ is an upper bound of $a_i, i=1,2,\ldots$.
Let $c\in {\mathcal L}_0$ be any other upper bound of $a_i,
i=1,2,\ldots$. Then $a_i\leq c$ for all $i$ implies that
$c',a_1,a_2,\ldots$ are mutually orthogonal. By (Oiii), there is
an element $d\in {\mathcal L}_0$ such that for every $p\in
\Sigma$,
$$
\mu_p(d)=\mu_p(c')+\sum_{i=1}^{\infty} \mu_p(a_i)=\mu_p(c')+\mu_p(a)
$$
>From this we obtain $\mu_p(c)=\mu_p(d')+\mu_p(a)$, which entails
by (i) that $a\leq c$.

Let $u$ be the supremum of $(a_i)_{i=1}^{\infty}$.
The we have $\forall p\in \Sigma$, $u\in \xi(p)$ if and only
if $a_i\in \xi(p)$ for some $i$. But then $u\in \xi(p)$ if and only
if $a\in \xi(p)$, which entails that $u=a$.
\end{proof}

We will say that a set ${\mathcal F}$ of functions $f:L\to [0,1]$
defined on a partially ordered set $L$ is {\it order determining}
if $a\leq b \, \Leftrightarrow \,  \forall  f\in \mathcal F,
f(a)\leq f(b)$.

\begin{theorem}\label{th:5} Let $({\mathcal L},\Sigma, \xi)$ be
a state-property system,  ${\mathcal L}_0\subseteq {\mathcal L}$,
and let ${\mathcal M}:=\{ \mu:\Sigma \times {\mathcal L}_0\to
[0,1]\}$ satisfy conditions {\rm (Oi)-(Oiii)}. Then the set $L_0$
is a $\sigma$-orthomodular poset and the set ${\mathcal M}$ is
order determining for $L_0$. Moreover, for every $a\in L_0$,
$a\neq 0$, there is $p\in \Sigma$ such that $\mu_p(a)=1$.
\end{theorem}
\begin{proof} By definition, the set ${\mathcal L}_0$ with the ordering
inherited from $\mathcal L$ is a partially ordered set. By Lemma
\ref{le:5}, ${\mathcal L}_0$ is an orthocomplemented set such that
the supremum of every pairwise orthogonal sequence exists in
${\mathcal L}_0$. Moreover, $\mathcal M$ is ordering for
${\mathcal L}_0$. Assume $a\leq b$, $a,b\in {\mathcal L}_0$. Then
$\forall p\in \Sigma$, $\mu_p(a)\leq \mu_p(b)$ implies
$\mu_p(a)+\mu_p(b')\leq 1$, so that $a\vee b'$ exists in
${\mathcal L}_0$ and $\mu_p(a\vee b')=\mu_p(a)+\mu_p(b')$ for all
$p\in \Sigma$, which entails that
$\mu_p(b)=\mu_p(a)+\mu_p(a'\wedge b)$ for all $p\in \Sigma$, hence
$\mu_p(b)=\mu_p(a\vee(a'\wedge b))$ for all $p\in \Sigma$, so by
Lemma \ref{le:5} (ii), $b=a\vee(a'\wedge b)$, which is the
orthomodular law. Hence ${\mathcal L}_0$ is a
$\sigma$-orthocomplete orthomodular poset.

Let $a\in {\mathcal L}$, $a\neq 0$, and assume that  $\forall p\in
\Sigma$, $a\notin \xi(p)$. Then  the implication
$$
\forall r\in \Sigma : a\in \xi(r) \, \Rightarrow \, 0\in \xi(r)
$$
holds, which by (5) of Definition \ref{de:1} means that $a=0$, a
contradiction. If $0\neq a\in {\mathcal L}_0$, then $a\in \xi(p)$
for at least one $p\in \Sigma$ means that $\mu_p(a)=1$.
\end{proof}

>From now on, we will write $(\Sigma, {\mathcal L}, {\mathcal L}_0,
\xi)$ to denote a state property system for which there is
${\mathcal L}_0\subset {\mathcal L}$  with a system of functions
$\mu_s, s\in \Sigma$ such that conditions (Oi), (Oii) and (Oiii)
are satisfied.

\begin{definition}\label{de:supp} Let $(\Sigma, {\mathcal L},
{\mathcal L}_0, \xi)$ be given.  We will say that $\mu_p$ has a
{\it support} {\rm (}in ${\mathcal L}_0${\rm )} if there is an
element $b\in {\mathcal L}$ {\rm (}$b\in {\mathcal L}_0${\rm )}
such that $\forall a\in {\mathcal L}_0$, $\mu_p(a)=1$ iff $b\leq
a$.
\end{definition}

Clearly, if a support exists, it is unique.

\begin{proposition}\label{pr:atom} Let $(\Sigma,{\mathcal L},
\xi)$ be a state property system, satisfying condition {\rm (A)}.
 For $p\in \Sigma$, let $a_p:=\bigwedge \{ a:a\in \xi(p)\}$.
Then $a_p, p\in \Sigma$, coincide with the atoms in $\mathcal L$.
Moreover, $a\in \xi(p)$ if and only if $a_p\leq a$.
\end{proposition}
\begin{proof}
Observe that condition {\rm(A)} is implies also condition
\begin{enumerate}
\item[{\rm(A')}] for all $p\in \Sigma$, the element $a_p=\bigwedge \{
a:a\in \xi(p)\}\neq 0$.
\end{enumerate}
Indeed, by Definition \ref{de:1}, $\bigwedge \{ a:a\in \xi(p)\}\in
\xi(p)$, and $0\not\in \xi(p)$. Hence $a_p\in \xi(p)$, and
clearly, $a_p$ is the smallest element in $\xi(p)$.  Assume that
$a_p\in \xi(r)$, $r\in \Sigma$. Now $a_p\leq a$ for all $a\in
\xi(p)$ implies that $a\in \xi(r)$ for all $a\in \xi(p)$, hence
$\xi(p)\subset \xi(r)$. By condition (A) then $p=r$.

Assume $b\leq a_p$, $b\neq 0$, then $\exists r, b\in \xi(r)$ and
we have
\begin{eqnarray*}
 &\forall r\in \Sigma, b\in \xi(r) \, \Rightarrow \,
a_p\in \xi(r)
\Rightarrow \, a\in \xi(r)\, \forall a\in \xi(p)\\
&\xi(p)\subset \xi(r)\Rightarrow p=r \Rightarrow b\in \xi(p)\,
\Rightarrow \, a_p\leq b.
\end{eqnarray*}
This proves that $a_p$ is an atom in ${\mathcal L}$.

Now let $a$ be an atom of $\mathcal L$. Then there is $r\in
\Sigma$ with $a\in \xi(r)$, hence $a_r\leq a$. Since $a_r$ is an
atom, $a_r=a$.
\end{proof}

Notice that under conditions of Proposition \ref{pr:atom}, the
element $a_p$ is a support of $\mu_p$.

\begin{theorem}\label{th:atomistic} Under the suppositions of
{\rm Proposition \ref{pr:atom}}, ${\mathcal L}$ is an atomistic
lattice.
\end{theorem}
\begin{proof} Let $b\in {\mathcal L}$, put $c=\bigvee \{ a_s: b\in
\xi(s)\}$. Then clearly $c\leq b$, and if $b\in \xi(p)$, then
$a_p\leq c$ implies $c\in \xi(p)$, therefore $b=c$.
\end{proof}

\begin{theorem}\label{th:orthocompl} Let $(\Sigma,{\mathcal
L}, {\mathcal L}_0,\xi)$ be a state property system satisfying
condition {\rm (A)} and
\begin{enumerate}
\item[{\rm(B)}] For every $s\in \Sigma$, $a_s$ belongs to
${\mathcal L}_0$.
\item[{\rm(C)}] For every $b\in {\mathcal L}$, $b=\bigwedge \{
a_s':b\leq a_s'\}$.
\end{enumerate}
Then ${\mathcal L}$ with the mapping $b':=\bigvee \{ a_s: b\leq
a_s'\}$ is a complete, atomistic, orthocomplemented lattice.
\end{theorem}
\begin{proof}  Owing to Theorem \ref{th:atomistic}, it suffices
to prove that ${\mathcal L}$ is orthocomplemented. (i) If $b\leq
c$, then $\{ a_s: c\leq a_s'\}\subset \{ a_s: b\leq a_s'\}$, which
by (C) implies $c'\leq b'$. (ii) From $b'\leq a_s'$ iff $a_s\leq
b$ we obtain that $(b')'=\bigvee \{a_s: b'\leq a_s'\} =\bigvee \{
a_s: a_s\leq b\}=b$. It remains to prove that $b\wedge b'=0$.
Assume that $a_s\leq b, a_s\leq b'$. By (i) and (ii), $a_s\leq
b\leq a_s'$, which contradicts (B), so $b\wedge b'=0$.  By duality
we get $b\vee b'=I$.
\end{proof}

\begin{definition}\label{de:3.3.14} Let
$(\Sigma, {\mathcal L}, {\mathcal L}_0, \xi)$ be given. We will
say that $p$ is orthogonal to $q$, $p,q\in \Sigma$, if there is
$a\in {\mathcal L}_0$ such that $\mu_p(a)=1$ and $\mu_q(a)=0$
(equivalently, $p\perp q$ if $a\in \xi(p)$, $a'\in \xi(q)$). If
$p$ is orthogonal to $q$ we will write $p\perp q$.
\end{definition}

It is obvious that the relation $\perp$ is symmetric and
anti-reflexive. For $T\subset \Sigma$, we put $T'=\{ p\in \Sigma
:p\perp T\}$, where $p\perp T$ means that $p\perp t$ for all $t\in
T$. Clearly, $\emptyset '=\Sigma$, $T\subseteq T''$ and
$T_1\subset T_2$ implies $T_1'\supset T_2'$ $\forall
T_1,T_2\subset \Sigma$. If $s,p\in \Sigma$ have supports in
${\mathcal L}_0$, then $s\perp p$ if and only if their supports
are orthogonal.

Denote by ${\bar T}^0$ the set of all $s\in \Sigma$ such that
$\forall a\in {\mathcal L}_0$, $a\in \xi(t)\, \forall t\in T\,
\Rightarrow \, a\in \xi(s)$. That is, ${\bar T}^0$ is the set of
all superpositions of $T\subset \Sigma$ with respect to ${\mathcal
L}_0$. Equivalently, ${\bar T}^0=\{ s\in \Sigma:\forall a\in
{\mathcal L}_0,\, a_t\leq a\, \Rightarrow\, a_s\leq a\}$. Clearly,
${\bar T}\subset {\bar T}^0$.

\begin{proposition}\label{pr:T} Let
$(\Sigma, {\mathcal L}, {\mathcal L}_0, \xi)$ be a state property
system satisfying conditions {\rm (A),(B),(C)}. Then for every
$T\subset \Sigma$, ${\bar T}={\bar T}^0$.
\end{proposition}
\begin{proof} It suffices to prove that ${\bar T}^0\subset {\bar
T}$. We have $s\in {\bar T}^0$ iff $\forall a\in {\mathcal L}_0$,
$a_t\leq a \, \forall t\in T \, \Rightarrow \, a_s\leq a$. Let us
take $b\in {\mathcal L}$, and assume that $a_t\leq b\, \forall
t\in T$. By property (C), $b=\bigwedge \{ a_r': b\leq a_r'\}$,
which yields $a_t\leq a_r'$ for all $t\in T$ and $r$ such that
$b\leq a_r'$.  From $s\in {\bar T}^0$ we obtain that $a_s\leq
a_r'$ for all corresponding $r$, and therefore $a_s\leq \bigwedge
\{ a_r': b\leq a_r'\}=b$. In other words, $\bigvee_{t\in T} \xi(t)
\subset \xi(s)$, hence $s\in {\bar T}$.
\end{proof}

\begin{proposition}\label{pr:Guz} Let $(\Sigma, {\mathcal L},
{\mathcal L}_0, \xi)$. Suppose that  {\rm (A), (B)} are satisfied.
Then for any $T\subset \Sigma$ we have $T''={\bar T}^0$.
\end{proposition}
\begin{proof} We follow the proof of \cite[Proposition
3.3.15]{PtPu}. We will identify $\mu_s$ with $s\in \Sigma$ and
write $T(a)=k$ if $\mu_t(a)=k \, \forall t\in T$.  First we show
that $T'=\emptyset$ if and only if $\{ a\in {\mathcal L}_0:
T(a)=1\}=\{ 1\}$. Assume that $T'=\emptyset$ and let $a\in
{\mathcal L}_0$ be such that $a\neq 1$ and $T(a)=1$. Since $a'\neq
0$, there is $p\in \Sigma$ such that $p(a')=1$. But then $p(a)=0$,
so that $a\in T'$, a contradiction. Now assume that $\{ a\in
{\mathcal L}_0: T(a)=1\}=\{ 1\}$ and also that $p\in T'$. Then for
the supports we have $a_p\perp a_t\,\,  \forall t\in T$. Hence
$t(a_p')=1$ for all $t\in T$, which is again a contradiction.

To prove the equality $T''={\bar T}^0$, assume first that
$T'=\emptyset$. We have already proved that then $\{ a\in
{\mathcal L}_0: T(a)=1\}=\{ 1\}$, which implies ${\bar T}^0=\Sigma
=T''$.

Assume that $T'\neq \emptyset$ and also that $p\in {\bar T}^0$. We
will show that $p\in T''$. Assume that $q\in T'$, then $a_q\perp
a_t \, \forall t\in T$, and hence $T(a_q')=1$. This implies
$a_q'\in \bigcap_{t\in T} \xi(t)$, which implies that $a_q'\in
\xi(p)$. This implies $q\perp p$, which implies that ${\bar
T}^0\subset T''$.

Assume that $p\in T''$ and also that $T(a)=1$ for some $a\in
{\mathcal L}_0$. Without loss of generality we may assume that
$a\neq 1$. We have $a_q\leq a'$ iff $q(a')=1$. But $q(a')=1$
implies that $q\in T'$. This means that $q\perp p$, and so
$a_p\perp a_q$ for all $q$ such that $q(a')=1$.  Hence $a_q\leq
a'$ implies $a_q\leq a_p'$, so that $a'\leq a_p'$, so that
$p(a)=1$. This shows that $p\in {\bar T}^0$ and this completes the
proof.
\end{proof}

As a corollary of Propositions \ref{pr:T} and \ref{pr:Guz}, we
obtain the following.
\newtheorem{corollary}{Corollary}

\begin{corollary}\label{co:F} Let $(\Sigma, {\mathcal L},
{\mathcal L}_0 \xi)$ be a state property system satisfying {\rm
(A),(B),(C)}. Then for every $T\subset \Sigma$, ${\bar T}=T''$.
\end{corollary}

\begin{theorem}\label{th:3.3.16} Let $(\Sigma,{\mathcal
L}, {\mathcal L}_0, \xi)$ be a state property system satisfying
{\rm (A),(B)}. Define ${\mathcal F}^0(\Sigma):=\{S\subset \Sigma
:S={\bar S}^0\}$. Then the mapping $S\mapsto S'$ is an
orthocomplementation on ${\mathcal F}^0(\Sigma)$. Consequently,
${\mathcal F}^0(\Sigma)$ is a complete, atomistic,
orthocomplemented lattice. If also {\rm (C)} holds, then $S\mapsto
S'$ is an orthocomplementation on ${\mathcal F}(\Sigma)$, and
${\mathcal F}(\Sigma)$ is a complete, atomistic, orthocomplemented
lattice.
\end{theorem}
\begin{proof} It is easy to check that $S\mapsto {\bar S}^0$ is
a closure operation, and hence ${\mathcal F}^0(\Sigma)$ is a
complete lattice with lattice operations $S\wedge T=S\cap T$ and
$S\vee T=(S\cup T)^{-0}$. Owing to property (A), ${\mathcal
F}^0(\Sigma)$ is atomistic. To prove orthocomplementation, observe
that $S\subset T\, \Rightarrow \, T'\subset S'$ and $S\wedge
S'=\emptyset$ follow directly from the definition of the mapping
$S\mapsto S'$. Property $S''=S$ for $S\in {\mathcal F}^0(\Sigma)$
follows from Proposition \ref{pr:Guz}. The remaining statement
follows from Corollary \ref{co:F}.
\end{proof}

\begin{definition}\label{de:cartan}
Suppose that $(\Sigma, {\mathcal L}, \xi)$ is a state property
system. The map $\kappa: {\mathcal L}\to {\mathcal P}(\Sigma)$
defined by
\begin{equation}\label{eq:kappa}
\kappa(a)=\{ p\in \Sigma: a\in \xi(p)\}
\end{equation}
is called the Cartan map.
\end{definition}

According to \cite[Proposition 5]{aertsetal01}, $\kappa :{\mathcal
L}\to (\kappa({\mathcal L}), \subset, \cap)$ has the following
properties:
\begin{eqnarray}
\kappa(1)&=&\Sigma,\\
\kappa(0)&=&\emptyset,\\
a\leq b&\Leftrightarrow& \kappa(a)\subset \kappa(b),\\
\kappa(\bigwedge_i a_i)=\bigcap_i \kappa(a_i).
\end{eqnarray}
That is, $\kappa$ is an isomorphism of complete lattices. Moreover,
by \cite[Theorem 2]{aertsetal01}, $\{ \kappa(a):a\in {\mathcal L}\}$
is an intersection system. Consequently, the operator $cl:Y\mapsto
\bigcap \{ \kappa (a):Y\subset \kappa(a)\}$ is a closure operator
\cite{FF}.

Next lemma shows that $\kappa(a)$ is closed under superpositions.

\begin{lemma}\label{le:kappa} Let $(\Sigma, {\mathcal L}, \xi)$ be
a state property system. For all $a\in {\mathcal L}$,
$\kappa(a)\in {\mathcal F}(\Sigma)$.
\end{lemma}
\begin{proof} For every $a\in {\mathcal L}$ we have
$\kappa(a)\subset  \kappa(a)^-$. Observe that $p\in \kappa(a)
\Leftrightarrow a\in \xi(p)$. Let $s\in {\bar \kappa(a)}$, then
$\bigcap_{p\in \kappa(a)}\xi(p)\subset \xi(s)$ implies $a\in
\xi(s)$, which means that $s\in \kappa(a)$.
\end{proof}

\begin{proposition}\label{pr:kappa}Let
$(\Sigma, {\mathcal L},\xi)$ be a state property system such that
condition {\rm (A)} is satisfied. Then $\kappa(\mathcal L)$ and
${\mathcal F}(\Sigma)$ are isomorphic as complete atomistic
lattices.
\end{proposition}
\begin{proof} By Lemma \ref{le:kappa}, the range of $\kappa$ is in
${\mathcal F}(\Sigma)$. By \cite[Proposition 5]{aertsetal01},
$\kappa(\mathcal L)$ and ${\mathcal F}(\Sigma)$ are isomorphic as
complete lattices. Let $a\in \mathcal L$ be an atom. By definition,
$\kappa(a)=\{ p\in \Sigma :a\in \xi(p)\}$. By (A), $a\in \xi(p)$ iff
$a_p\leq a$, hence $a_p=a$ because $a$ is an atom. By (A) we may
conclude that $\kappa(a)=\{ p\}$.
\end{proof}

\begin{theorem}\label{th:kappa}  Let $(\Sigma,{\mathcal
L},{\mathcal L}_0,\xi)$ be a state property system satisfying {\rm
(A),(B)}. Then the mapping $\kappa :{\mathcal L}_0\to {\mathcal
F}(\Sigma)$, $a\mapsto \kappa(a)$ has the following properties:
\begin{enumerate}
\item[{(\rm(i)}] If $a\wedge b$ exists in ${\mathcal L}_0$, then
$\kappa(a\wedge b)=\kappa(a)\wedge \kappa(b)$.
\item[{\rm(ii)}] For all $a\in {\mathcal L}_0$,
$\kappa(a')=\kappa(a)'$.
\end{enumerate}
Consequently, $\kappa({\mathcal L}_0)$ and ${\mathcal L}_0$ are
isomorphic as atomistic $\sigma$-orthomodular posets.

If also condition (C) is satisfied, then $\kappa(\mathcal L)$ and
$\mathcal F(\Sigma)$ are isomorphic as complete, atomistic
orthocomplemented lattices.
\end{theorem}
\begin{proof} (i) Suppose that $a\wedge b$ exists in ${\mathcal L}_0$.
Obviously, $\kappa(a\wedge b)\leq \kappa(a)\wedge \kappa(b)$.
Suppose that $s\in \kappa(a)\wedge \kappa(b)= \kappa(a)\cap
\kappa(b)$. This gives $a,b\in \xi(s)$, hence $a_s\leq a, a_s\leq
b$, consequently $a_s\leq a\wedge b$, i.e. $s\in \kappa(a\wedge
b)$.

(ii) Assume that $p\in \kappa(a)'$, where $a\in {\mathcal L}_0$
with $0< a <1$. Then $p\perp q$ for all $q\in \kappa(a)$. It
follows that $\forall q\in \kappa(a),\, a_q\leq a_p'$. Hence
$a_p\leq (\vee_{q\in \kappa(a)}a_q)'=a'$. This proves
$\kappa(a)'\leq \kappa(a')$.

Now let $p\in \kappa(a')$, then $a_p\leq a'=(\vee \{a_q: a_q\leq
a\})'$, hence $a_p\leq a_q'$ for all $q\in \kappa(a)$, which
entails $p\in \kappa(a)'$.

The rest follows by Proposition \ref{pr:kappa}.
\end{proof}

\section{Superposition principle and sectors}
Let $(\mathcal L, \Sigma, \xi)$ be a state property system such
that  property (A) and 3-MSP are satisfied. By Theorem
\ref{th:1}, $(\Sigma, *)$, where $p*s=\lambda \{p,s\}=\{ p,s\}^-$
is a projective geometry.

\begin{definition}\label{de:SP} We will say that a superposition
principle {\rm (SP}, for short{\rm )} is satisfied in $(\mathcal
L, \Sigma, \xi)$, if for every $p,q\in \Sigma$, $p\neq q$, there
is $r\in \{ p,q\}^-$ such that $r\neq p, r\neq q$.
\end{definition}

The following statement is straightforward.

\begin{theorem}\label{th:ired} Let $(\mathcal L, \Sigma, \xi)$ be
a state property system such that {\rm (A), 3-MSP}and {\rm SP}
are satisfied. The $(\Sigma, *)$ is an irreducible projective
geometry.
\end{theorem}

The notion of a sector was introduced in \cite{Pu} (see also
\cite[Definition 3.2.7]{PtPu}). Roughly speaking, a sector is a
maximal $\lambda$-closed subset of $\Sigma$ in which SP holds.

\begin{definition}\label{de:sect} A nonempty subset $S\subset \Sigma$
is called a sector if the following conditions hold:
\begin{enumerate}
\item[{\rm (i)}] $S\in {\mathcal L}(\Sigma)$;
\item[{\rm(ii)}] for any two different $p,q\in S$ we can find
$r\in \{ p,q\}^-$ distinct from $p$ and $q$;
\item[{\rm(iii)}] if $q\in \Sigma \setminus S$, then  $\{ p,q\}^-=\{ p,q\}$
for every $p\in S$.
\end{enumerate}
\end{definition}

A basic property of sectors is the following.
\begin{lemma}\label{le:sect} If $S,P$ are sectors, then either
$S=P$ or $S\cap P=\emptyset$.
\end{lemma}
\begin{proof} Assume that $S\neq P$. Then there is $q\in S\setminus
P$ (or $q\in P\setminus S$), and by (ii) of Definition
\ref{de:sect}, $\{ s,q\}^-\neq \{ s,q\}$ whenever $s\in S\cap P$, while
by (iii) of Definition \ref{de:sect}, $\{ s,q\}^-=\{ s,q\}$. This
contradiction implies that $S\cap P=\emptyset$.
\end{proof}

\begin{theorem}\label{th:sect}{\rm \cite{PtPu}} Let
$(\mathcal L, \Sigma, \xi)$ be a state property system such that
{\rm (A)} and {\rm 3-MSP} are satisfied. Then $\Sigma$ can be
written as a set theoretical union of sectors.
\end{theorem}
\begin{proof} Let us define a binary relation $\approx$ on
$\Sigma$ as follows:   (i) for every $s\in \Sigma$, $s\approx s$,
(ii) for distinct $s,t\in \Sigma$,  $s\approx t$ if there is $r\in
\{ s,t\}^-$, $r\neq s, r\neq t$. We will prove that $\approx$ is
an equivalence relation. Reflexivity and symmetry are clear from
the definition. To prove transitivity, assume that $p\approx r$
and $r\approx s$. With no loss of generality, we may assume that
$p,r,s$ are mutually different. Let $x\in \{ p,r\}^-\setminus \{
p,r\}$, $y\in \{ r,s\}^-\setminus \{ r,s\}$. By 2-MSP we have $\{
p,r\}^-=\{ p,x\}^-=\{ r,x\}^-$, $\{ r,s\}^-=\{r,y\}^-=\{ s,y\}^-$.
Moreover,  $r\in \{ x,p\}^-$ implies $y\in \{ x,p,s\}^-=\lambda
\{x,p,s\} \subset S$ by 3-MSP. If $y\in \{ p,s\}^-$ and $y\neq p$,
then $y$ is a minimal superposition of $\{ p,s\}$, and hence
$p\approx s$. If $y=p$, then $p\in \{ r,s\}^-$ implies $r\in \{
s,p\}^-$, hence $p\approx s$. If $y=x$, then $\{ r,x\}^-=\{
r,y\}^-$ implies $\{ p,r\}^-=\{ r,s\}^-$, $p\in \{ r,s\}^-$, hence
$r\in \{ s,p\}^-$ and $p\approx s$. Finally, if $y$ is a minimal
superposition, then $\{ y,x\}^-\cap \{p,s\}^-\neq \emptyset$
implies that $p\approx s$.

Let ${\hat s}$ denote the equivalence class containing $s\in
\Sigma$. We may write $\Sigma =\bigcup \{ {\hat s}:s\in \Sigma\}$.
It can be easily seen that $\hat s$ is a sector for every $s\in
\Sigma$.
\end{proof}

Sectors can be characterized by the closure operator $\lambda$ as
follows.

\begin{theorem}\label{th:dirk} Let $(\mathcal L, \Sigma, \xi)$ be
a state property system such that {\rm (A)} and {\rm 3-MSP} are
satisfied. Let $\Sigma =\bigcup_i S_i$, where $S_i\in {\mathcal
L}(\Sigma)$ and {\rm (SP)} is satisfied on $S_i$, $\forall i$.
Then $S_i$ are sectors if and only if they are $\lambda$-clopen
sets.
\end{theorem}
\begin{proof} By Theorem \ref{th:sect}, $\Sigma$ can covered by
sectors, which are $\lambda$-closed. Let $S$ be a sector. To prove
that $S$ is clopen, it suffices to prove that $\Sigma \setminus S$
is $\lambda$-closed. Assume that $p,q\in \Sigma \setminus S$ and
let $r$ be a minimal superposition of $p,q$. If $r\in S$, then by
2-MSP, $p\in \lambda \{ r,q\}$. Since $r\in S$ and $q\notin S$,
and $S$ is a sector, we have $\lambda \{r,q\}= \{r,q\}$, which is
a contradiction. Therefore $r\in \Sigma \setminus S$. This proves
that sectors are $\lambda$-clopen sets.

Conversely, let $S$ be a $\lambda$-clopen set such that
SP is satisfied on $S$. Then conditions (i)
and (ii)  of Definition \ref{de:sect} are satisfied.  To
prove (iii), assume that $p\in S, q\notin S$, and $r\in \{
p,q\}^-$, $r\neq p,q$, then either $r\in S$ or $r\notin S$. If
$r\in S$, we get $q\in \{ r,p\}^-$, which contradicts the
supposition that $S$ is $\lambda$-closed. If $r\notin S$, we get
$p\in \{ r,q\}^-$, which contradicts the supposition that $S$ is
open. It follows that $\{ p,q\}^-=\{ p,q\}$, hence $S$ is a
sector.
\end{proof}

\begin{definition}\label{de:super} We say that an element $a\in
\mathcal L$ is classical (or a superselection rule) if there is an
element $a'\in \mathcal L$ such that for every $s\in \Sigma$,
$a\in \xi(s) \Leftrightarrow  a'\notin \xi(s)$.
\end{definition}

Clearly, $0$ and $1$ are classical elements.

\begin{theorem}\label{th: super} Let $a\in \mathcal L$ be a
classical element. Then $\kappa(a)=\{ s\in \Sigma: a\in \xi(s)\}$
is a clopen set in ${\mathcal F}(\Sigma)$.
\end{theorem}
\begin{proof} We have $\Sigma =\{ s:a\in \xi(s)\} \cup \{ s: a'\in
\xi(s)\}$. By symmetry, it suffices to prove that $S:=\{ s: a\in
\xi(s)\}$ belongs to ${\mathcal F}(\Sigma)$. It easily follows
from the fact that $r\in {\bar S}$ iff $\bigcap\{ \xi(s):s\in S\}
\subset \xi(r)$, which entails that  if $a\in \xi(s)\, \forall
s\in S$, then $a\in \xi(r)$, hence $r\in S$.
\end{proof}
>From the fact that ${\mathcal F}(\Sigma)\subset {\mathcal
L}(\Sigma)$, we obtain that $\kappa(a)$ is clopen also in
${\mathcal L}(\Sigma)$.

We recall that an element $z$ in a lattice $L$ with $0$ and $1$ is
{\it central} when there exist two lattices $L_1$ and $L_2$ and an
isomorphism between $L$ and the direct product $L_1\times L_2$
such that $z$ corresponds to the element $(1_1,0_2)\in L_1\times
L_2$.  (cf e.g. \cite[Definition (4.12)]{MM}. Evidently $0$ and
$1$ are central elements.

\begin{lemma}\label{le:MM} {\rm \cite[Theorem(4.13)]{MM}} An
element $z$ of a lattice $L$ with $0$ and $1$ is central if and
only if there is an element $z'$ in $L$ such that
\begin{equation}\label{eq:center} a=(a\wedge z)\vee(a\wedge z')
=(a\vee z)\wedge(a\vee z')\ \mbox{for every}\  a\in L.
\end{equation}

If $L$ is orthocomplemented, then $z$ is central if and only if
the first equality in {\rm(\ref{eq:center})} is satisfied for
every $a\in L$ {\rm(\cite[Lemma (29.9)]{MM})}.
\end{lemma}

\begin{theorem}\label{th:centclas} Let $({\mathcal L}, {\mathcal
L}_0, \Sigma,\xi)$ be a state-property system such that conditions
{\rm (A), (B), (C)} are satisfied. Then an element $c\in {\mathcal
L}$ is central if and only if $c$ is classical.
\end{theorem}
\begin{proof} If properties (A), (B), (C) are satisfied,
then ${\mathcal L}$ is a complete, atomistic, orthocomplemented
lattice, and $\kappa :{\mathcal L}\to {\mathcal F}(\Sigma)$ is an
isomorphism (Theorem \ref{th:kappa}).

Let $c$ be a central element of $L$, then by (\ref{eq:center}),
for every atom $a\in {\mathcal L}$, $a=(a\wedge c)\vee(a\wedge
c')$, hence either $a=a\wedge c$, or $a=a\wedge c'$. By
Proposition \ref{pr:kappa}, $\kappa(a)=\{ s\}$ for some $s\in
\Sigma$. Moreover, $\kappa(a)=\kappa(a\wedge c)\vee \kappa(a\wedge
c')$, hence either $\kappa(a\wedge c)=\{ s\}$, or $\kappa(a\wedge
c')=\{ s\}$, that is, either $c\in \xi(s)$ or $c'\in \xi(s)$. This
entails that $c$ is classical.

Conversely, if $c$ is classical, i.e., for every $s\in \Sigma$,
either $s\in \kappa(c)$ or $s\in \kappa(c')$, then for every
$a\in {\mathcal L}$,
\begin{eqnarray*}
 a&=&\bigvee \{ a_s:s\in \kappa(a)\}\\
 &=& \bigvee
\{ a_s: s\in \kappa(a)\cap \kappa(c)\} \vee \bigvee \{ a_s: s\in
\kappa(a)
\cap \kappa(c')\}\\
&=& \bigvee \{ a_s:s\in \kappa(a\wedge c)\}\vee \bigvee \{ a_s:
s\in \kappa(a\wedge c')\},
\end{eqnarray*}
and consequently, $a=(a\wedge c)\vee(a\wedge c')$. By Lemma
\ref{le:MM}, $c$ is central element of ${\mathcal L}$.
\end{proof}

\section{Closed subspaces and Mackey property}
Throughout this section we will use the following notations:
\begin{equation}\label{eq:subsp}
\mbox{For any}\, A,B\in {\mathcal L}(\Sigma),\, A\sqcup
B:=\lambda(A\cup B).
\end{equation}

\begin{equation}\label{eq:clsubsp} \mbox{For any}\,  A,B\in
{\mathcal F}(\Sigma),\, A\vee B:=(A\cup B)^-.
\end{equation}

For infima in both ${\mathcal L}(\Sigma)$, ${\mathcal F}(\Sigma)$
we use the same notation $A\wedge B(=A\cap B)$.

In \cite{FF}, the following definitions were introduced, and the
equivalence of the following three categories was proved.

\begin{definition}\label{de:MG} A Mackey geometry is a projective
geometry $G$ together with a subset $\mathcal S$ of subspaces of
$G$ satisfying the following axioms:
\begin{enumerate}
\item[{\rm(i)}] ${\mathcal A}\subseteq {\mathcal S}$ implies
$\bigcap {\mathcal A}\in {\mathcal S}$ (hence ${\mathcal S}$ is an
intersection system),
\item[{\rm(ii)}] $\emptyset \in {\mathcal S}$,
\item[{\rm(iii)}] if $E\in {\mathcal S}$, then $a\vee E\in
{\mathcal S}$ for every $a\in G$.
\end{enumerate}
The elements of ${\mathcal S}$ are called the closed subspaces of
$G$. An isomorphism of Mackey geometries is an isomorphism of
projective geometries $g:G_1\rightarrow G_2$ satisfying
$S\in{\mathcal S}_1$ iff $g(E)\in {\mathcal S}_2$ (where $E$ is
any subspace of $G_1$).
\end{definition}

\begin{definition}\label{de:ML} A Mackey lattice is a projective
lattice $L$ together with an operator $x\rightarrow c(x)$
satisfying the following axioms:
\begin{enumerate}
\item[{\rm (i)}] $x\leq c(x)$ for every $x\in L$,
\item[{\rm(ii)}] $x\leq c(y)$ implies $c(x)\leq c(y)$,
\item[{\rm(iii)}] $c(0)=0$,
\item[{\rm(iv)}] if $x=c(x)$, the $a\vee x=c(a\vee x)$ for every
atom $a$ in $L$.
\end{enumerate}
An element $x\in L$ is closed if $x=c(x)$. An isomorphism of
Mackey lattices is an isomorphism of (projective) lattices
$h:L_1\rightarrow L_2$ satisfying $h(c_1(x))=c_2(h(x))$ for every
element $x\in L_1$.
\end{definition}

For any lattice $L$ we shall denote by $A_L$ the set of all atoms
of $L$. We say that a lattice $L$ has the {\it intersection
property} (cf \cite[Definition 2.5.1]{FF}) if one has
\begin{equation}\label{eq:ip}
a,b\in A_L, a\neq b, x\in L \ \mbox{and}\ a\leq b\vee x\
\Rightarrow \ \exists c\in A_L \ \mbox{with}\ c\leq (a\vee
b)\wedge x.
\end{equation}
If $L$ is an atomistic lattice, the following conditions are
equivalent:
\begin{enumerate}
\item $L$ is upper and lower semimodular\footnote{A lattice $L$ is called
{\it upper semimodular} if $u\wedge v\lessdot v$ implies $u\lessdot
u\vee v$, and $L$ is {\it lower semimodular} if $u\lessdot u\vee v$
implies $u\wedge v\lessdot v$. Here $a\lessdot b$ means that $b$
covers $a$. },
\item $L$ has the covering property\footnote{A lattice $L$ has the
{\it covering property} if for $x\in L$ and any atom $a\in L$ one
has, $a\wedge x=0\ \implies \ x\lessdot a\vee x$.}
\item $L$ has the intersection property.
\end{enumerate}
Moreover, the implications $1\Rightarrow 2\Rightarrow 3$ hold for
any lattice.

\begin{definition}\label{de:IL} An intersection lattice is a
complete atomistic lattice  $C$ having the intersection property.
(Equivalently,  $C$ is both upper and lower semimodular.)
\end{definition}

Let $L_1$ and $L_2$ be Mackey lattices. We say that a morphism
$f:L_1\to L_2$ is {\it continuous} if
\begin{equation}\label{eq:cont}
f(c_1(x))\leq c_2(f(x)) \, \mbox{for every} \, x\in L_1.
\end{equation}

\begin{theorem}\label{th:13.3.8}{\rm \cite[Theorem 13.3.8]{FF}}
The categories of Mackey geometries, of Mackey lattices and of
intersection lattices are equivalent. This means that one has a
functor $\mathcal L$ from Mackey geometries to Mackey lattices, a
functor $\mathcal C$ from Mackey lattices to to intersection
lattices, a functor $\mathcal G$ from intersection lattices to
Mackey geometries, and natural isomorphisms $G\cong {\mathcal
G}({\mathcal C}({\mathcal L}(G)))$, $L\cong {\mathcal L}({\mathcal
G}({\mathcal C}(L)))$ and $C\cong {\mathcal C}({\mathcal
L}({\mathcal G}(C)))$.
\end{theorem}

In our setting, we obtain the following result.

\begin{theorem}\label{th:Mackey} Let $({\mathcal L},
{\mathcal L}_0, \Sigma, \xi)$ satisfy properties {\rm (A), (B),
(C)} and {\rm 3-MSP}. Then ${\mathcal L}(\Sigma)$ with the closure
operation $c(A)={\bar A}$, $A\in {\mathcal L}(\Sigma)$, is a
Mackey lattice.
\end{theorem}
\begin{proof}(cf \cite[Proposition 3.3.18]{PtPu}).
Properties (i) -(iii) of Definition
\ref{de:ML} are clear. We have to prove only property (iv).

By Theorem \ref{th:3.3.16}, ${\mathcal F}(\Sigma)$ is a complete,
atomistic, orthocomplemented lattice with the orthocomplementation
$S\mapsto S'$. We will use the fact that ${\mathcal
L}(\Sigma)\supset {\mathcal F}(\Sigma)$,  and ${\mathcal
L}(\Sigma)$ is modular (Theorem \ref{th:2}). We must show that if
$S\in {\mathcal F}(\Sigma)$ and $p\in \Sigma \setminus S$, $p\vee
S=p\sqcup S$. Modularity of ${\mathcal L}(\Sigma)$ implies that
$S\prec S\sqcup p$ (that is, $S\sqcup p$ covers $S$). Dually,
$(S\sqcup p)'=S'\wedge \{ p\}'\prec S'$. Then there is an atom
$q\in \Sigma$ such that $(S\sqcup \{ p\})'\sqcup\{ q\}=S'$. Then
$((S\sqcup \{ p\})'\sqcup\{ q\})'=S''=S\prec (S\sqcup \{ p\})''$.
This entails $(S\sqcup \{ p\})''=S\vee \{ p\}=S\sqcup \{ p\}$.
\end{proof}

\begin{corollary}\label{co:3-f}
If $({\mathcal L}, {\mathcal L}_0, \Sigma, \xi)$ satisfy
properties {\rm (A), (B), (C)} then for every $S\in {\mathcal
F}(\Sigma)$ and a finite dimensional element $P=\{ p_1,\ldots,
p_n\}^-$, we have $S\vee P=S\sqcup P$.
\end{corollary}

In accordance with theorems \ref{th:13.3.8} and \ref{th:Mackey},
if a state property system $({\mathcal L},\Sigma,\xi)$ satisfies
conditions (A), (B), (C) and 3-MSP, we may consider $\Sigma$ with
elements of ${\mathcal F}$ as closed subspaces as  Mackey
geometry, ${\mathcal L}(\Sigma)$ with the operator $S\to {\bar S}$
as a Mackey lattice, and   ${\mathcal F}(\Sigma)$) as intersection
lattice. Indeed, by theorem \ref{th:Mackey}, ${\mathcal
L}(\Sigma)$ with the operation $S\mapsto {\bar S}$ is a Mackey
lattice. By \cite[Proposition 13.2.7]{FF}, the set ${\mathcal
F}(\Sigma)$ is an intersection lattice for the induced order. The
infimum of any subset ${\mathcal A}\subset {\mathcal F}(\Sigma)$ is the
element $\wedge {\mathcal A}$ and the supremum is $\vee {\mathcal
A}=(\sqcup {\mathcal A})^-$. Moreover, the atoms of ${\mathcal
F}(\Sigma)$ are the atoms of ${\mathcal L}(\Sigma)$, that is,
elements of $\Sigma$ . Further, ${\mathcal F}(\Sigma)$ being an
intersection lattice, the set of all atoms $\Sigma$ of ${\mathcal
F}(\Sigma)$  is a projective geometry (cf \cite[2.5.7]{FF} and
Theorem \ref{th:1}), and the set of closure subspaces coincides
with the sets  $\{ F\subset \Sigma: F\in {\mathcal F}(\Sigma)\}$ as
closed subspaces. Owing the isomorphism between ${\mathcal
F}(\Sigma)$ and $\mathcal L$, the lattice ${\mathcal L}$ can be
considered as an intersection lattice with the atoms $\{ a_s\in L:
s\in \Sigma \}$.

In the sequel, we will need the following definition.

\begin{definition}\label{de:13.4.6} {\rm \cite[Definition 13.4.6]{FF}}
A Mackey lattice $L$ is called regular if for every closed element
$x\in L$ and every atom $a\not\leq x$, there exists a closed
coatom $h\in L$ such that $x\leq h$ and $a\not\leq h$.
\end{definition}

\section{Orthogeometries, ortholattices and orthosystems}

\begin{definition}\label{de:orthogeom} {\rm \cite[Definition
14.1.1]{FF}} An orthogeometry is a projective geometry with a
relation $\perp$, called orthogonality, which satisfies the
following axioms:
\begin{enumerate}
\item{\rm $(O_1)$} $a\perp b$ implies $b\perp a$,
\item{\rm $(O_2)$} if $a\perp p$, $b\perp p$ and $c\in a*b$, then
$c\perp p$,
\item{\rm $(O_3)$} if $a,b,c\in G$ and $b\neq c$, then there is $p\in
b*c$ with $p\perp a$,
\item{\rm $(O_4)$} for every $a\in G$ there exists $b\in G$ with
$a\not\perp b$.
\end{enumerate}

An isomorphism of orthogeometries is an isomorphism of projective
geometries $g:G_1\to G_2$ satisfying $a\perp b$ iff $ga\perp gb$.
\end{definition}

For any subset $A\subseteq G$ the {\it orthogonal set}
$A^{\perp}:=\{ x\in G: x\perp a \ \mbox{for every}\ a\in A\}$ is a
subspace of $G$ by condition $(O_2)$. A point $a$ of an
orthogeometry $G$ is called a {\it null point} if $a\in
a^{\perp}$. The geometry is called {\it non-null} if it contains a
non-null point and {\it pure} if every point is non-null
\cite[Definition 14.1.7]{FF}.

\begin{definition}\label{de:ortholat} An ortholattice\footnote{Please do not mistake it with
orthocomplemented lattice, which is sometimes also called
ortholattice.} is a projective lattice together with an operator
$x\mapsto x^{\perp}$ which satisfies the following conditions:
\begin{enumerate}
\item $x\leq x^{\perp \perp}$ for every $x\in L$,
\item $x\leq y$ implies $y^{\perp}\leq x^{\perp}$,
\item $0^{\perp \perp} =0$,
\item if $x=x^{\perp \perp}$, then $a\vee x=(a\vee x)^{\perp \perp}$
for every atom $a\in L$.
\end{enumerate}

An isomorphism of ortholattices is an isomorphism of lattices
$f:L_1\to L_2$ such that $f(x^{\perp})=(fx)^{\perp}$ for every
element $x\in L_1$.
\end{definition}

\begin{proposition}\label{pr:14.2.4} If $L$ is an ortholattice,
then $L$ together with the operator $c(x):=x^{\perp \perp}$ is a
regular Mackey lattice.
\end{proposition}

\begin{definition}\label{de:orthosyst} An orthosystem is an
intersection lattice $C$ together with an operator $x\mapsto x'$
satisfying the following conditions:
\begin{enumerate}
\item $x=x''$ for every $x\in C$,
\item $x\leq y$ implies $y'\leq x'$.
\end{enumerate}
An isomorphism of orthosystems is an isomorphism of lattices
$h:C_1\to C_2$ such that $h(x')=(hx)'$ for every $x\in C_1$.
\end{definition}

By \cite[Remark 14.2.7]{FF}, instead of an intersection lattice it
is enough to require that $C$ is a complete atomistic lattice
satisfying the exchange property.

In \cite{FF}, it is proved that there is a triple correspondence
between orthogeometries and ortholattices and orthosystems
\cite[Proposition 14.2.11]{FF}. We summarize the results in the
next theorem.

\begin{theorem}\label{th:ortho}
\begin{enumerate}
\item If $L$ is an orthogeometry, then the projective lattice
${\mathcal L}(G)$ together with the operator $E\mapsto E^{\perp}$
is an ortholattice {\rm \cite[Proposition 14.2.5]{FF}}.

\item Let $L$ be an ortholattice. Denote by ${\mathcal C}(L)$ the
set of all closed element $x=x^{\perp \perp}$ of $L$.  Then
${\mathcal C}(L)$ together with the operator $x\mapsto x^{\perp}$
is an orthosystem  (for the induced order) {\rm \cite[Proposition
14.2.8]{FF}}.

\item Let $C$ be an orthosystem. Then the projective geometry
${\mathcal G}(C)$ consisting of the set $A_C$ of all atoms of $C$
and the operator $*$, $a*b=\{ c\in A_C:c\leq a\vee b\}$,  together
with the relation $a\perp b$ iff $a\leq b'$, is an orthogeometry
\cite[proposition 14.2.9]{FF}.
\end{enumerate}
\end{theorem}

\begin{theorem}\label{th:orthosyst} Let
$({\mathcal L},{\mathcal L}_0,\Sigma,\xi)$ satisfy properties {\rm
(A), (B), (C)} and {\rm 3-MSP}. Then ${\mathcal F}(\Sigma)$ is an
orthosystem.
\end{theorem}
\begin{proof} By Theorems \ref{th:Mackey} and \ref{th:13.3.8},
${\mathcal F}(\Sigma)$ is an intersection lattice. By Theorem
\ref{th:3.3.16}, the mapping $S\mapsto S'$ is an
orthocomplementation on ${\mathcal F}(\Sigma)$, which implies the
desired result.
\end{proof}

\newtheorem{remark}{Remark}
\begin{remark}\label{re:regular} If $({\mathcal L},{\mathcal L}_0,\Sigma,\xi)$
satisfies properties {\rm (A), (B), (C)} and {\rm 3-MSP} then, according
to {\rm Theorem \ref{th:ortho}}, $(\Sigma, \lambda)$ is an
orthogeometry and ${\mathcal L}(\Sigma)$ with the closed subspaces
${\mathcal F}(\Sigma)$ is an ortholattice. Moreover, since ${\bar
S}=S''$, in accordance with {\it \cite[Proposition 14.2.4]{FF}},
${\mathcal F}(\Sigma)$ is a regular Mackey lattice.
\end{remark}

\section{Representations in vector spaces}

Let $V$ be any vector space over a field $K$. We emphasize that
the dimension of $V$ is arbitrary (possibly infinite)
and $K$ is allowed to be a skew field (often called division ring).

\begin{proposition}\label{pr:vect}{\rm \cite[Proposition 2.1.6]{FF}}.
Let $V$ be any vector space.
On $V^*:=V\setminus \{ 0\}$ one
defines a binary relation as follows: $x\sim y$ iff $x,y$
if $x,y$ are linearly dependent. Since this is an equivalence relation,
the quotient set ${\mathcal P}(V):=V^*/\sim$ is well defined
and becomes a projective geometry if for any elements
$X,Y,Z\in {\mathcal P}(V)$ one defines $\ell(X,Y,Z)$ iff
$X,Y,Z$ have linearly dependent representatives $x,y,z$.
\end{proposition}

\begin{theorem}\label{th:baer} Let $G$ be an irreducible
projective geometry containing at least four independent points.
Then there exists a (left) vector space $V$ over a field $K$
such that $G$ is isomorphic to ${\mathcal P}(V)$.
\end{theorem}

\begin{definition}\label{de:hermit} {\rm \cite[Definition 14.1.5]{FF}}
Let $V$ be a vector space over $K$.
A map:$\Phi:V\times V\to K$ is called a reflexive (or also symmetric)
sesquilinear form if there exists an anti-isomorphism of fields
$\sigma :K\to K$ such that the following axioms are satisfied:
\begin{enumerate}
\item [{\rm(1)}] $\Phi(x_1+x_2,y)=\Phi(x_1,y)+\Phi(x_2,y)$ and
$\Phi(\lambda x,y)=\lambda .\Phi(x,y)$,
\item[{\rm(2)}] $\Phi(x,y_1+y_2)=\Phi(x,y_1)+\Phi(x,y_2)$ and
$\Phi(x,\mu y)=\Phi(x,y).\sigma(\mu)$;
\item[{\rm(4)}] $\Phi(x,y)=0$ iff $\Phi(y,x)=0$.
\end{enumerate}
A map $\Phi :V\times V\to K$ is called a Hermitian
form if there exists an involution $\sigma :K\to K$, i.e.,
an anti-isomorphism of order 2, such that the following axioms
are satisfied:
\begin{enumerate}
\item[{\rm(1)}] $\Phi(x-1+x_2,y)=\Phi(x_1,y)+\phi(x_2,y)$ and
$Phi(\lambda x,y)=\lambda .\Phi(x,y)$,
\item[{\rm(4)}] $\Phi(x,y)=\sigma(\Phi(y,x))$ for all $x,y\in V$.
\end{enumerate}
Obviously, these two axioms imply both {\rm (2)} and {\rm (3)}.
Finally, we recall that the form $\Phi$ is non-singular if
$\Phi(x,y)=0$ for all $y\in V$ implies $x=0$.
\end{definition}

\begin{proposition}\label{pr:14.1.6}{\rm \cite[Proposition 14.1.6]{FF}}
If $\Phi :V\times V\to K$ is a non-singular reflexive sesquilinear form,
then the projective geometry ${\mathcal P}(V)$ together with the relation
$\perp$ defined by $[x]\perp [y]$ iff $\Phi(x,y)=0$ is an orthogeometry.
\end{proposition}

\begin{definition}\label{de:nullp} A point $a$ of an orthogeometry $G$
is called a null point if $a\in a^{\perp}$. The orthogeometry is called
non-null if it contains a non-null point and pure if every point is non-null.
\end{definition}

Let $V$ be a pre-Hilbertian space over $\mathbb R, \mathbb C, \mathbb H$,
then trivially the associated orthogeometry ${\mathcal P}(V)$ is pure.

\begin{theorem}\label{th:14.1.8}{\rm \cite[Theorem 14.1.8]{FF}}
Let $V$ be a vector space of dimension $\geq$ 3 over a field $K$,
and suppose that ${\mathcal P}(V)$ together with the relation
$\perp$ is an orthogeometry. Then there exists a non-singular reflexive
sesquilinear form $\Phi :V\times V\to K$ which induces the orthogonality
$\perp$ in the sense of Proposition {\rm \ref{pr:14.1.6}}. Moreover, if
${\mathcal P}(V)$ is non-null, then $\perp$ can be induced by a
(non-singular) Hermitian form.
\end{theorem}

We call states $S:=\{ s_1,\ldots,s_n\}$  in $\Sigma$ {\it independent} if  $\forall i$,
$s_i\notin \lambda(S\setminus s_i)$.

\begin{theorem}\label{th:premain}
Let $({\mathcal L}, \Sigma, \xi)$ be a state property system such that
conditions {\rm (A), SP, 3-MSP} are satisfied. Assume that there exist at
least four independent states in $\Sigma$. Then there is a field $K$ and
a vector space $V$ over $K$ such that the set ${\mathcal L}(\Sigma)$ of
all linear subspaces of $\Sigma$ is isomorphic to the lattice
${\mathcal L}(V)$ of all linear subspaces of $V$.
\end{theorem}

\begin{theorem}\label{th:main} Let $({\mathcal L}, \Sigma, \xi)$ be
a state property system such that
conditions {\rm (A), (B),(C), SP, 3-MSP} are satisfied.
Assume that there exist at
least four independent states in $\Sigma$. Then there exists a field $K$,
an involutive ant-automorphism $^*:K\to K$, a vector space $V$ over $K$
and a Hermitian form $f:V\times V\to K$ such that ${\mathcal F}(\Sigma)$
is orthoisomorphic to the set ${\mathcal L}_f(V)$ of all closed subspaces
of $V$.
\end{theorem}

(See \cite{MM} for the ideas of proof).

\end{document}